\documentclass[preprint,pra,showpacs,nofootinbib]{revtex4}
\usepackage{graphicx}
\usepackage{float,epsfig}
\usepackage{dcolumn}
\usepackage{bm}
\usepackage{graphicx}
\usepackage{dcolumn}
\usepackage{bm}
\usepackage{amsmath,amssymb,amsthm}
\usepackage[colorlinks=true,linkcolor=blue]{hyperref}
\begin{document}
\title{\bf Entanglement generation in atoms immersed in a thermal bath of external quantum scalar fields with a boundary}
\author{Jialin Zhang }
\affiliation{Department of Physics and Institute of  Physics,\\
Hunan Normal University, Changsha, Hunan 410081, China}
\author{Hongwei Yu\footnote{Corresponding author}}
\affiliation{ CCAST(World Lab.), P. O. Box 8730, Beijing, 100080, P.
R. China and Department of Physics and Institute of  Physics,\\
Hunan Normal University, Changsha, Hunan 410081, China
\footnote{Mailing address}}


\begin{abstract}

We examine the entanglement creation between  two mutually
independent two-level atoms immersed in a thermal bath of quantum
scalar fields in the presence of  a perfectly reflecting plane
boundary. With the help of the master equation that describes the
evolution in time of the atom subsystem obtained, in the
weak-coupling limit, by tracing over environment (scalar fields)
degrees of freedom, we find that the presence of the boundary may
play a significant role in the entanglement creation in some
circumstances and the new parameter, the distance of the atoms from
the boundary, besides the bath temperature and the separation
between the atoms, gives us more freedom in manipulating
entanglement generation. Remarkably, the final remaining
entanglement in the equilibrium state is independent of the presence
of the  boundary.

\end{abstract}

\maketitle

\section{Introduction}
Quantum entanglement has now been recognized as a key resource in
quantum information science~\cite{information}, since it plays a
primary role in quantum communication~\cite{Bennett}, quantum
teleportation~\cite{telportation}, quantum
cryptography~\cite{cryptography} and so on. An interesting issue
in the discussions for the essence of  entanglement, which has
attracted a lot of attention, is the relationship between
entanglement and environment.  It is known that an environment
usually leads to decoherence and noise, which may cause
entanglement that might have been created before to disappear.
However, in certain circumstances, the environment may enhance
entanglement rather than destroying
it~\cite{pr1,pr2,pr3,pr4,pr5,pr6}. The reason is that an external
environment can also provide an indirect interaction between
otherwise totally uncoupled subsystems through correlations that
exist. For example, correlations in vacuum fluctuations or
fluctuations at finite temperature can provide such an
interaction, when entanglement generation is considered in systems
in external quantum fields.

Recently Benatti et al have discussed, in the framework of open
systems, entanglement generation
 for two, independent uniformly accelerating
two-level atoms interacting with a set of scalar fields in vacuum.
In the weak coupling limit,  the completely positive dynamics for
the atoms as a subsystem has been derived by tracing over the
field degrees of freedom~\cite{Benatti1}, and there it has been
shown that the asymptotic equilibrium state of the atoms turns out
to be entangled even if the initial state is separable. Similar
results have been obtained by considering  two atoms immersed in a
thermal bath of scalar particles at a finite
temperature~\cite{Benatti2}, where, in contrast to
Ref.~\cite{Benatti1}, two atoms are assumed to be at a finite
separation. It is found that for any fixed, finite separation,
there always exists a temperature below which entanglement
generation occurs as soon as time starts to become nonzero and for
the vanishing separation the entanglement thus generated persists
even in the late-time asymptotic equilibrium state. Therefore, one
can manipulate the entanglement production by controlling two
controllable parameters: the bath temperature and the separation
of the atoms.

In the above studies, the field correlation functions that
characterize the fluctuations of fields play a very important role
in determining whether entanglement is generated. On the other
hand, it is well-known that the presence of boundaries in a flat
spacetime  modifies the  fluctuations of quantum fields, and it
has been demonstrated that this modification  can lead to a lot of
novel effects, such as the Casimir effect~\cite{cas}, the
light-cone fluctuations when gravity is quantized~\cite{YU}, the
Brownian (random) motion of test particles in an electromagnetic
vacuum~\cite{yu2}, and the modification for the radiative
properties of uniformly accelerated atoms~\cite{yu3}.

 A question then arises naturally as to what happens to the
entanglement generation if the field correlations are modified by
the presence of a reflecting boundary. Now we have one more
controllable parameter other than the separation and the bath
temperature, i.e., the distance of the atoms from the boundary and
another interesting question is what is the role that the new
parameter plays in the entanglement generation. These are
questions we are going to address in the present paper.  We shall
examine the entanglement generation of two non-interacting
two-level atoms immersed in a thermal bath of scalar particles
subjected to a perfectly reflecting plane boundary.  With the help
of the master equation that describes the evolution of the open
system (atoms plus external thermal fields) in time, we find that
the presence of the boundary may play an significant  role in
controlling the entanglement creation in some circumstances and
the new parameter, the distance of the atoms from the boundary,
gives one more freedom in controlling the entanglement generation.
It is, however, interesting that the probable remaining
entanglement for the asymptotic  equilibrium state at late times
is not dependent on the presence of the boundary.


\section{Two Atom Master Equation}

The system we shall examined is composed of two independent
two-level atoms in weak interaction with a set of massless quantum
scalar fields at a finite temperature $T$. We assume that a
perfectly reflecting plane boundary for the scalar fields is
located at $z=0$ in space and one atom is placed at point ${\bf
x}_1$ and the other at ${\bf x}_2$.  Without loss of generality,
we take the total Hamiltonian to have the form
\begin{equation}\label{H}
 H=H_s+H_\phi+
\lambda\;H'\;.
\end{equation}
Here $H_s$ is the Hamiltonian of the two atoms,
 \begin{equation}
H_s=H_S^{(1)}+H_s^{(2)},\ \ H_s^{(\alpha)}={\omega\over 2}\,
n_i\,\sigma_i^{(\alpha)}, \quad(\alpha=1,2), \ \ \label{H-atom}
 \end{equation}
where $\sigma_i^{(1)}=\sigma_i\otimes{\sigma_0},\ \
\sigma_i^{(2)}={\sigma_0}\otimes\sigma_i$,  $\sigma_i,(i=1,2,3)$
are the Pauli matrices, $\sigma_0$  the $2\times2$ unit matrix,
$\mathbf{n} =(n_1,n_2,n_3)$  a unit vector, $\omega$ the energy
level spacing, and summation over repeated index is implied.
$H_\phi$ is the standard Hamiltonian of massless, free scalar
fields, details of which is not relevant here and $H'$ is the
 Hamiltonian that describes the interaction between the two atoms with the external scalar
fields which is assumed to be weak. The general form for $H'$  can
be written as
\begin{equation}
H'=\sum_{\mu=0}^{3}\,[(\sigma_{\mu} \otimes \sigma_0)\Phi_{\mu}(t,
{\bf x}_1)+(\sigma_{0} \otimes \sigma_{\mu})\Phi_{\mu}(t, {\bf
x}_2)\,]\;.
\end{equation}
Now we assume that the scalar fields can be expanded as
\begin{equation}
\Phi_{\mu}(x)=\sum^N_{a=1}\,[\chi_\mu^a\phi^{(-)}(x) +
(\chi_\mu^a)^*\phi^{(+)}(x)]\;,
\end{equation}
where $\phi^{(\pm)}(x)$ are positive and negative energy field
operators  of the massless scalar field, and $\chi_\mu^a$ are
complex coefficients that "embed" the field modes into the
two-dimensional detector Hilbert space and play the role of
generalized coupling constants~\cite{Benatti1}.
 It
should be pointed out that the coupling constant $\lambda$ in
(\ref{H}) is small, and this is consistent with the assumption that
the interaction of the atom with the scalar fields is weak.

 It is well-known that the evolution of the total system  density (i.e., the two atoms plus the
environment) in time obeys the Liouville equation
$\partial_t\rho_{tot}(t)=- i [H, \rho_{tot}(t)] $ with the initial
total density having a generic form
$\rho_{tot}(0)=\rho(0)\otimes\rho_B$,  where  the environment
fields are taken to be in a thermal state characterized by
$\rho_B$ and the atom in an initial state $\rho(0)$. Since our
interest is in the dynamics for the two atoms only,  we must trace
over the environment degrees of freedom and concentrate on the
analysis of the reduced time evolution,
$\rho(t)=Tr_\phi[\rho_{tot}]$. Provided that the field
correlations decay sufficiently fast at large time separations, or
much faster than the characteristic evolution time of the
subsystem alone, the reduced density of the two-atom subsystem can
be proven, in the limit of weak-coupling, to obey an equation in
the Kossakowski-Lindblad form~\cite{Lindblad,Benatti1, Benatti2,
pr5}
\begin{equation}
{\partial\rho(t)\over \partial t}= -i \big[H_{\rm eff},\,
\rho(t)\big]
 + {\cal L}[\rho(t)]\ ,
\label{master}
\end{equation}
with
\begin{equation}
H_{\rm eff}=H_S-\frac{i}{2}\sum_{\alpha,\beta=1}^2
H_{ij}^{(\alpha\beta)}\ \sigma_i^{(\alpha)}\,\sigma_j^{(\beta)}\ ,
\label{10}
\end{equation}
and
\begin{equation}
{\cal L}[\rho]={1\over2} \sum_{\alpha,\beta=1}^2
C_{ij}^{(\alpha\beta)}\big[2\,
\sigma_j^{(\beta)}\rho\,\sigma_i^{(\alpha)}
-\sigma_i^{(\alpha)}\sigma_j^{(\beta)}\, \rho
-\rho\,\sigma_i^{(\alpha)}\sigma_j^{(\beta)}\big]\ .
\label{lindblad}
\end{equation}
The coefficients of the matrix $C_{ij}^{(\alpha\beta)}$ and
$H_{ij}^{(\alpha\beta)}$ are determined by the field correlation
functions in the thermal state $\rho_\beta$:
\begin{equation}
\mathrm{}G_{ij}^{\alpha\beta}(t-t')={}_\beta\langle\Phi_i(t,\mathbf{x}_{\alpha})\Phi_j(t',\mathbf{x}_{\beta})
\rangle_\beta\;.\label{green}
\end{equation}
 The corresponding Fourier and Hilbert transforms read respectively
\begin{equation}
{\cal G}_{ij}^{(\alpha\beta)}(\lambda)=\int_{-\infty}^{\infty} dt \,
e^{i{\lambda}t}\, G_{ij}^{(\alpha\beta)}(t)\; , \label{fourierG}
\end{equation}
\begin{equation}
{\cal K}_{ij}^{(\alpha\beta)}(\lambda)=\int_{-\infty}^{\infty} dt \,
{\rm sign}(t)\, e^{i{\lambda}t}\, G_{ij}^{(\alpha\beta)}(t)=
\frac{P}{\pi i}\int_{-\infty}^{\infty} d\omega\ \frac{ {\cal
G}_{ij}^{(\alpha\beta)}(\omega) }{\omega-\lambda} \;, \label{kij}
\end{equation}
where $P$ denotes principal value. One can show that the
Kossakowski matrix $C_{ij}^{(\alpha\beta)}$ can be written
explicitly as
\begin{equation}
C_{ij}^{(\alpha\beta)}=\sum_{\xi=+,-,0} {\cal
G}_{kl}^{(\alpha\beta)}(\xi\omega)\, \psi_{ki}^{(\xi)}\,
\psi_{lj}^{(-\xi)}\; , \label{cij}
\end{equation}
where
\begin{equation}
\psi_{ij}^{(0)}=n_i\, n_j\ ,\qquad \psi_{ij}^{(\pm)}={1\over
2}\big(\delta_{ij} - n_i\, n_j\pm i\epsilon_{ijk} n_k\big)\ .
\label{psij}
\end{equation} Similarly, the coefficients of $H_{ij}^{\alpha\beta}$
can be obtained by replacing ${\cal
G}_{kl}^{(\alpha\beta)}(\xi\omega)$ with ${\cal
K}_{kl}^{(\alpha\beta)}(\xi\omega)$ in the above expressions. For
the sake of simplicity of our treatment, we now assume that the
field correlation functions are diagonal such that
\begin{equation}
\mathrm{}G_{ij}^{\alpha\beta}(t-t')=\delta_{ij}G (t-t', {\bf
x}_\alpha-{\bf x}_\beta) \;.
\end{equation}
This requirement can be fulfilled by demanding that  the coupling
coefficients $\chi^a_\mu$ satisfy the following condition
 \begin{equation}
\sum^N_{a=1}\,\chi^a_\mu(\chi^a_\nu)^*=\delta_{\mu\nu}
\end{equation}
or by assuming that the field components $\Phi_i (x)$ are
independent.

\section{The condition for entanglement creation}

With the basic formalism established, now we shall start to
examine whether entanglement can be generated between two
independent atoms in external thermal fields at a finite
temperature $T$ subjected to a reflecting boundary (i.e. fields
are constrained to vanish on the boundary), in particular, what is
the influence the presence of a boundary that modifies the quantum
correlations of the fields will have on entanglement generation.

\begin{figure}[htbp]
\centering
\includegraphics[scale=1]{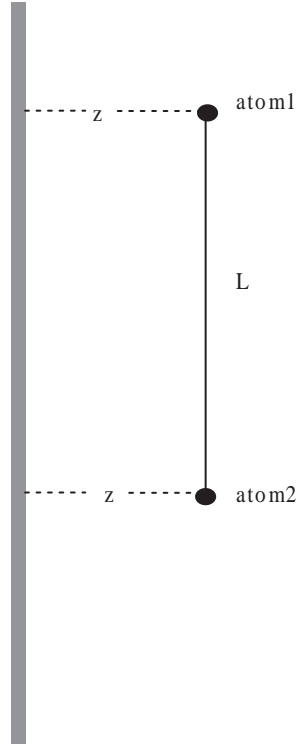}
\caption{The infinite conducting  plate  is  taken to lie along the
plane $z=0$, the distance between each atom and the plate is $z$,
the atom separation is $L$  }\label{tu1}
\end{figure}

For the sake of simplicity, let us further assume that  two atoms
are separated from each other by a distance $L$ and are at an
equal distance $z$ from the boundary(Fig.~(\ref{tu1})), i.e.,
$z_1=z_2=z$. Due to the assumption that the fields reflect from
the boundary completely, we can use the  method of images
~\cite{greenf} to find the field correlation functions (
Eq.~(\ref{green}) ),

  \begin{eqnarray}
G_{ij}^{(11)}(t-t')=G_{ij}^{(22)}(t-t')=&&-{1\over4\pi^2}\sum_{m=-\infty}^{\infty}
\bigg[{\delta_{ij}\over(t-t'-im\beta-i\epsilon)^2}\nonumber\\
 &&-{\delta_{ij}\over(t-t'-im\beta-i\epsilon)^2-(2z)^2}\bigg]\;,\label{g11}
\end{eqnarray}
\begin{eqnarray}
G_{ij}^{(21)}(t-t')=G_{ij}^{(12)}(t-t')=&&-{1\over4\pi^2}\sum_{m=-\infty}^{\infty}
  \bigg[{\delta_{ij}\over(t-t'-im\beta-i\epsilon)^2-L^2}\nonumber\\&&
-{\delta_{ij}\over(t-t'-im\beta-i\epsilon)^2-(2z)^2-L^2}\bigg]\;,\label{g12}
\end{eqnarray}
where  $\beta=1/(k T)$. Plugging Eq.~(\ref{g11}) and Eq.~(\ref{g12})
into the Eq.~(\ref{fourierG}), we can easily obtain
\begin{eqnarray}
\nonumber &&{\cal G}^{(11)}_{ij}(\lambda)={\cal
G}^{(22)}_{ij}(\lambda)=\frac{\delta_{ij}}{2\pi}
\frac{\lambda}{1-e^{-\beta \lambda}}
-\frac{\delta_{ij}}{2\pi} \frac{\lambda}{1-e^{-\beta \lambda}}\frac{\sin(2 z\lambda)}{2z\lambda},\\
 &&{\cal G}^{(12)}_{ij}(\lambda)={\cal
G}^{(21)}_{ij}(\lambda)=\frac{\delta_{ij}}{2\pi}
\frac{\lambda}{1-e^{-\beta \lambda}}\ \frac{\sin(L \lambda)}{L
\lambda} -\frac{\delta_{ij}}{2\pi} \frac{\lambda}{1-e^{-\beta
\lambda}}\frac{\sin(\sqrt{L^2+4z^2}\lambda) }{\sqrt{L^2+4z^2}\lambda
}\; .\label{green-p}
\end{eqnarray}
According to Eq.~(\ref{cij}), we can write
\begin{eqnarray}
C_{ij}^{(11)}=C_{ij}^{(22)}=A_1 \delta_{ij}- iB_1
\epsilon_{ijk}n_k+C_1n_in_j\nonumber\\
C_{ij}^{(12)}=C_{ij}^{(21)}=A_2\delta_{ij}- iB_2
\epsilon_{ijk}n_k+C_2n_in_j\;,
\end{eqnarray}
and the corresponding  coefficients are
\begin{eqnarray}
&&A_1={\omega\over4\pi}{1+e^{-\beta{\omega}}\over1-e^{-\beta{\omega}}}\bigg[1-{\sin(2z\omega)\over2z\omega}\bigg]\;,\hspace*{10pt}
A_2={\omega\over4\pi}{1+e^{-\beta{\omega}}\over1-e^{-\beta{\omega}}}\bigg[{\sin(L\omega)\over{L\omega}}
-{\sin(\sqrt{L^2+4z^2}\omega)\over\sqrt{L^2+4z^2}\omega}\bigg]\;,
\nonumber\\
&&B_1={\omega\over4\pi}\bigg[1-{\sin(2z\omega)\over2z\omega}\bigg]\;,\hspace*{50pt}
B_2={\omega\over4\pi}\bigg[{\sin(L\omega)\over{L\omega}}-{\sin(\sqrt{L^2+4z^2}\omega)
\over\sqrt{L^2+4z^2}\omega}\bigg]\;,
\nonumber\\
&&C_1={\omega\over4\pi}{1+e^{-\beta{\omega}}\over1-e^{-\beta{\omega}}}\bigg[-1+{\sin(2z\omega)\over2z\omega}\bigg]\;,\nonumber\\
&&C_2={\omega\over4\pi}{1+e^{-\beta{\omega}}\over1-e^{-\beta{\omega}}}\bigg[-{\sin(L\omega)\over{L\omega}}+{\sin(\sqrt{L^2+4z^2}\omega)
\over\sqrt{L^2+4z^2}\omega}\bigg]\;.\label{abc}
\end{eqnarray}
 Similarly, the ${\cal K}_{ij}^{\alpha\beta}$ for the Hamiltonian $H_{\rm eff}$ can
 be obtained easily, but here we do not give the formulae in
 detail. As has already been discussed in detail elsewhere~\cite{Benatti1,Benatti2},
the effective Hamiltonian $H_{\rm eff}$ can be expressed as a sum
of three pieces. The first two correspond to the corrections of
the Lamb shift at a finite temperature which should be regularized
according to the standard procedures in quantum field theory and
nevertheless they can be accounted for by replacing $\omega$ in
the atom's Hamiltonian $H_S$ with a renormalized energy level
spacing
\begin{equation}
\tilde{ \omega} =\omega +i [{\cal K}^{11}(-\omega)-{\cal
K}^{11}(\omega)]\;.
\end{equation}
Meanwhile the third is an environment generated direct coupling
between the atoms and it is temperature independent. So the term
associated with $H_{\rm eff}$ in (\ref{master}) can be ignored,
since we are interested in the temperature-induced effects.
Henceforth, we will only study the effects produced
 by the dissipative part ${\cal L}[\rho(t)]$.

 Using the explicit form of the master equation~(\ref{master}), we can
 investigate the time evolution of the reduced density  matrix and  figure out whether the
 state of the two-level atom system is  an entangled one or not with the help of partial
 transposition  criterion~\cite{ppt}: a two-atom state $\rho(t)$ is entangled at $t$ if and only if
 the operation of partial transposition of $\rho(t)$ does not preserve its positivity.
 In general, the two-atom system in the thermal bath
will be subjected to decoherence and dissipation, which may
counteract the entanglement production, so that the final
equilibrium state is very likely to be separable (however this may
not always be true as we will demonstrate later).  But if we
consider the system evolving in a finite time, during which the
decoherence and dissipation are not dominant, the initial
separable state may evolve to an entangled one.  Here, we adopt a
simple strategy for ascertaining the entanglement creation at a
neighborhood of the initial time $t=0$, which has been introduced
in Ref.~\cite{pr5}.
 For simplicity, we let the initial pure, separable two-atom
 state be
 $\rho(0)=|+\rangle\langle+|\otimes|-\rangle\langle-|$ and consider
 the quantity
 \begin{equation}
{\cal Q}(t)=\langle\chi\vert\, \tilde{\rho}(t)\, \vert\chi\rangle\ ,
\end{equation}
where the tilde signifies partial transposition and $|\chi\rangle$
is a properly chosen $4$-dimensional vector. According to the
results of  Ref.~{\cite{Benatti2,pr5}),   entanglement is created
at the neighborhood of time $t=0$ (i.e., $\partial_t{\cal Q}(0)<
0$), if and only if
\begin{equation}\label{condition}
\langle{u}|C^{(11)}|u\rangle\langle{v}|(C^{(22)})^T)|v\rangle<|\langle{u}|Re(C^{(12)})|v\rangle|^2\;,
\end{equation}
 where the subscript $T$ means matrix
transposition and the three-dimensional vectors $|u\rangle$ and
$|v\rangle$ can be chosen in a simple form as
$u_i=v_i=\{1,-i,0\}$. Using  Eq.~(\ref{abc}),  we can calculate
Eq.~(\ref{condition}) for the vector $\mathbf{n}$ along the third
axis directly and deduce that the condition (\ref{condition})
becomes
\begin{equation}\label{bds}
\bigg({A_2\over {A_1}}\bigg)^2+\bigg({B_1\over {A_1}}\bigg)^2>1\;,
\end{equation}
where
\begin{equation}
\bigg({B_1\over
{A_1}}\bigg)^2=\bigg({1-e^{-\beta{\omega}}\over1+e^{-\beta{\omega}}}\bigg)^2\;,
\end{equation}
\begin{eqnarray}\label{a2a1}
\bigg({A_2\over
{A_1}}\bigg)^2&=&\bigg({\sin(L\omega)\over{L\omega}}-{\sin(\omega\sqrt{L^2+4z^2})\over{\omega\sqrt{L^2+4z^2}}}\bigg)^2/
\bigg(1-{\sin(2\omega{z})\over2z{\omega}}\bigg)^2\;.
\end{eqnarray}
For a given energy gap for the atoms,  ${B_1}^2/ {A_1}^2$ takes
the values in the interval [\,0,\,1\,] and is only
temperature-dependent, while the value of  ${A_2}^2/ {A_1}^2$ is
determined by two parameters, $z$ and $L$ and is temperature
independent. One can see that when the  temperature is zero i.e.
$\beta\rightarrow\infty$ and  $L$ is not infinite, the
inequality~(\ref{bds}) is always satisfied, therefore entanglement
is generated. At the same time, if the separation is vanishing
($L=0$), then ${A_2}^2/ {A_1}^2$ becomes unity and the
inequality~(\ref{bds}) is always obeyed and thus entanglement
created too, no matter where the atoms are placed, as long as the
bath temperature is not infinite.

Let us now discuss what happens when $z\rightarrow 0$ or $z\ll L$,
and $z\rightarrow \infty$ or $z\gg L$, i.e., when the atoms are
placed very close to and very far from the boundary. Expansion of
Eq.~(\ref{a2a1}) in power series of $z/L$ yields
\begin{eqnarray}\label{z0}
\bigg({A_2\over {A_1}}\bigg)^2&\approx&
\frac{9}{\omega^6L^6}\bigg[-\omega{L}\cos(\omega{L})+\sin(\omega{L})\bigg]^2+
\frac{18}{5\omega^6L^6}\bigg[\omega{L}\cos(\omega{L})-\sin(\omega{L})\bigg]\times\nonumber\\
&& \quad \bigg[\omega{L}(-15+\omega^2L^2)\cos(\omega{L})
+3(5-2\omega^2L^2)\sin(\omega{L})\bigg]\bigg({z\over L}\bigg)^2\;.
\end{eqnarray}
The corresponding form for $z/L\rightarrow \infty $ reads
\begin{eqnarray}\label{zd}
\bigg({A_2\over
{A_1}}\bigg)^2\approx
\frac{\sin^2(\omega{L})}{\omega^2L^2}+\frac{L}{z}\frac{\sin(\omega{L})}{\omega^3L^3}\bigg[\sin(\omega{L})\sin(2{\omega}z)-\omega{L}
\sin(\omega{L}\sqrt{1+4z^2/L^2})\bigg]\;,
\end{eqnarray}
where ${\sin^2(\omega{L})}/{\omega^2L^2}$ is just the value of
${A_2}^2/ {A_1}^2$ without the presence of the
boundary~\cite{Benatti2} (i.e., the corresponding value of
${A_2}^2/ {A_1}^2$ in the limit of $z\rightarrow\infty$). It is
interesting to note that in the limit of  $z/L\rightarrow 0$, the
leading term of ${A_2}^2/ {A_1}^2$ is independent on $z$ and is
only a function of $\omega L$ (refer to Eq.~(\ref{z0})). This
leading term differs from the value of ${A_2}^2/ {A_1}^2$ in the
case without the boundary. As a result, when the atoms are placed
very close to the boundary, the presence of the boundary will have
a significant effect in determining whether the
inequality~(\ref{bds}) is satisfied or whether entanglement is
created.  In Fig.~(\ref{sl}), the leading term of ${A_2}^2/
{A_1}^2$ when  $z/L\rightarrow 0$ is plotted as a function of
$\omega{L}$ vs ${A_2}^2/ {A_1}^2$ in the case without the
boundary. This Figure reveals that when $\omega{L}$ is small,
approximately smaller than 3, that is when the separation, $L$, is
approximately less than three times the characteristic wavelength
of the atom's radiation (but L is still large enough to maintain
$z\ll L$), the value of ${A_2}^2/ {A_1}^2$ in the case with the
presence of a boundary will be appreciably larger than that
without as long as $L$ is not vanishingly small. This means that
at a certain temperature the presence of a boundary would make the
atoms be entangled which otherwise still be separable. Therefore
the presence of the boundary provides us more freedom in
controlling entanglement creation in this case. However, when
$\omega{L}$ is large, i.e., the separation is much larger than the
characteristic wavelength of the atom's radiation, the value of
${A_2}^2/ {A_1}^2$ with the presence of the boundary generally
becomes smaller than  that without, since it decreases faster (as
power of $(\omega{L})^{-6}$ as opposed to $(\omega{L})^{-2}$) as
$\omega{L}$ grows. Therefore, in this case the presence of the
boundary will make the atoms less likely to be entangled than
otherwise. Meanwhile  when $z/L$ is very large, i.e., when the
atoms are very far from the boundary,
 the influence of the presence
of the boundary on the entanglement generation is negligible as
expected and this can be easily seen from Eq.~(\ref{zd}) since now
the leading term is the same as the value of ${A_2}^2/ {A_1}^2$ in
the unbounded case.

\begin{figure}[htbp]
\centering
\includegraphics[scale=1]{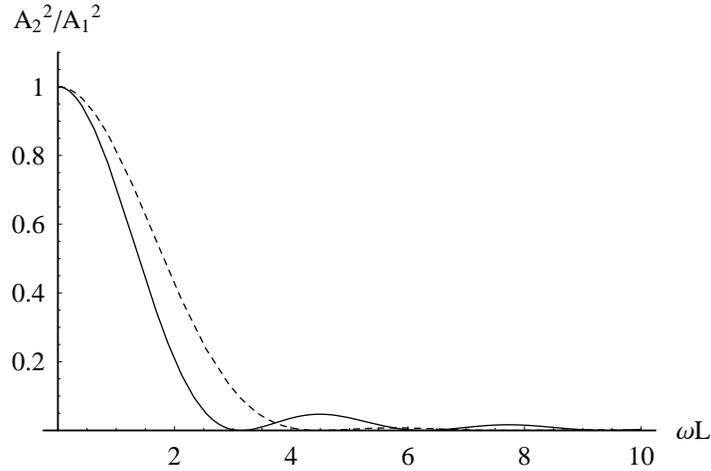}
\caption{The dashed line represents $A_2^2/A_1^2$ in the  limit of
$z/L\rightarrow 0$, and the solid line denotes the function,
$\sin^2(\omega{L})/(\omega{L})^2$, i.e, $A_2^2/A_1^2$ without
presence of a boundary.  Approximately in the interval [0,3] of
$\omega{L}$, the value of ${A_2}^2/ {A_1}^2$ with the presence of
the boundary is always larger than that without.} \label{sl}
\end{figure}


\begin{figure}[htbp]
\centering
\includegraphics[scale=1]{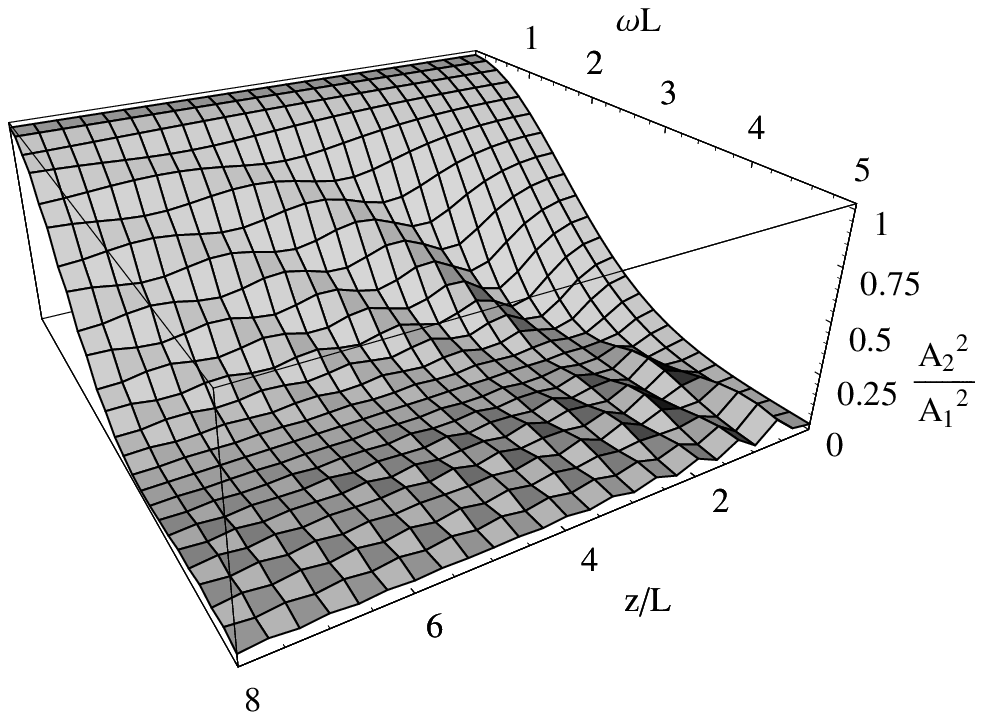}
\caption{ ${A_2}^2/ {A_1}^2$  as a function of  two dimensionless
variables, $z/L$ and $\omega{L}$. }\label{3d}
\end{figure}

To have a better understanding, we also plot, in Fig.~(\ref{3d}),
${A_2}^2/ {A_1}^2$ as a function of two dimensionless variables,
$z/L$ and $\omega{L}$,  according to Eq.~(\ref{a2a1}). One can see
from this figure that, as $z/L$ varies, appreciable oscillations
occur when $\omega{L}$  is of order one and  when $\omega{L}$ is
very small, the value of ${A_2}^2/ {A_1}^2$ is very close to unity
and does not oscillate significantly as $z/L$ varies. At same
time, ${A_2}^2/ {A_1}^2$ also decays very fast with the increase
of $\omega{L}$ and the oscillations (as $z/L$ varies) is damped
dramatically. So we conclude that both when $\omega{L}$ is very
small or very large, the variation of location of the atoms has no
significant influence on the entanglement generation. Note,
however,  that this by no means suggests that the presence of the
boundary does not affect the entanglement generation (refer to the
discussions in the preceding paragraph).

The next question we want to ask is what is the maximum difference
between the value of ${A_2}^2/ {A_1}^2$ with the presence of a
boundary and that without. We will try to answer the question
numerically and approximately.
\begin{figure}[htbp]
\centering
\includegraphics[scale=1]{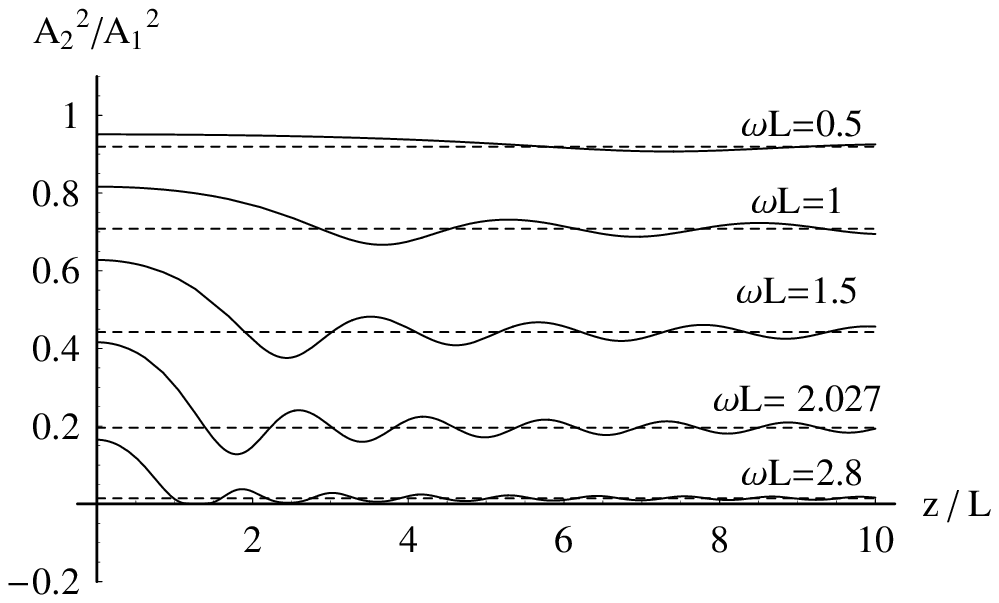}
\caption{The ratio ${A_2}^2/ {A_1}^2$ is described as a function of
the $z/L$ in the real line, with the parameter are selected
$\omega{L}=(0.5,1,1.5,2.027,2.8)$. The dashing line expresses the
value of $\sin^2(\omega{L})/(\omega{L})^2$ which is the value of
${A_2}^2/ {A_1}^2$ without the boundary conditions. As we can easily
see that all the values of ${A_2}^2/ {A_1}^2$ vibrate around the
$\sin^2(\omega{L})/(\omega{L})^2$.}\label{sz}
\end{figure}
For this purpose, let us plot, in Fig.~(\ref{sz}),  ${A_2}^2/
{A_1}^2$  as a function of $z/L$  with a set of fixed values of
the dimensionless parameter $\omega{L}$ in both the cases with and
without a boundary. Our numerical calculations as shown
illustratively in Fig.~(\ref{sz}) indicate that the maximum
fluctuation of ${A_2}^2/ {A_1}^2$ in the case with the presence of
the boundary around that without as a function of $z/L$ is at the
neighborhood of $\omega{L}\approx2.027$. Setting
$\omega{L}\approx2.027$, we find that the function of ${A_2}^2/
{A_1}^2$ vibrates around
$\sin^2(\omega{L})/(\omega{L})^2\approx0.196$. Due to the fact
that the swing of vibrating function ${A_2}^2/ {A_1}^2$ is slowly
decreasing with the increasing $z/L$, we can find out the maximum
value of ${A_2}^2/ {A_1}^2$, by adjusting the parameter $z/L$, to
be approximately $0.416$ achieved at $z/L\ll1$. This gives the
maximum  effects of the presence of the boundary on ${A_2}^2/
{A_1}^2$ or equivalently on entanglement creation.

To get a more concrete picture, let us take a typical transition
frequency of a hydrogen atom, $\omega \sim 10^{14} Hz$,  for an
example.  Then  $\omega{L}\approx2.027$ means
$L\approx6.08\times10^{-6} m$ which is much larger than the usual
size of an atom. It is easy to find that in the unbounded space the
inequality~(\ref{bds}) is satisfied or entanglement is created
between two atoms if the temperature is below $262.663$K. However,
with the presence of a boundary, we find that the upper bound in
temperature for entanglement generation can be increased to
$T<379.731$K. This is a hundred Kelvins improvement. Note for
$L\approx6.08\times10^{-6} m$, we still have a plenty of room to
satisfy $z/L\ll 1$, so the maximum value of $0.416$ is used for
${A_2}^2/ {A_1}^2$ here.

Finally, let us briefly discuss what happens if the two-atom
system is not aligned strictly parallel to the plane boundary.
Take the distance from the plane of the atom which is closer as
$z$, then the distance of the other atom from the plane will be be
larger or smaller than $z$ depending on whether the system is
inclined away from or towards the boundary. Therefore the effect
of inclination of the system is that the field correlation
function with respect to the atom which is displaced and cross
correlation function $G_{ij}^{(21)}(t-t')=G_{ij}^{(12)}(t-t')$
effectively get a smaller effective $z$ if the system is inclined
towards the plane and a larger effective one if otherwise (refer
to Eqs.~(\ref{g11},\ref{g12})).  Consequently, taking into account
the fact that  ${A_2}^2/ {A_1}^2$ is an oscillating function of
$z$ when the system is parallelly placed, one would expect that if
the two atom system is originally located parallel to the plane at
where this function is at its peak value the inclination in either
direction will make the entanglement creation less likely to
occur. In contrast, if the system is located at where this
function is at its local minimum the inclination in either
direction will make the entanglement generation more likely to
happen. However, when the system is placed at any point in the
interval where function ${A_2}^2/ {A_1}^2$ is monotonically
increasing, the atoms will be less likely to entangle if  the
inclination is towards the plane and more likely if otherwise.
Similarly, when the system is located at any point in the interval
where  function ${A_2}^2/ {A_1}^2$ is monotonically decreasing,
the entanglement generation will be more likely to come about if
the inclination is towards the plane and less likely if otherwise.

\section{The entanglement of the  equilibrium state with the boundary }

In the preceding Section, we find that, in certain circumstances,
the presence of a boundary plays a significant role in generating
entanglement between atoms initially prepared in a separable state
in a thermal bath of external quantum scalar fields and in fact
entanglement is created as soon as time starts if the
inequality~(\ref{bds}) is satisfied.  However, the
condition~(\ref{bds}) does not tell us whether the entanglement
thus generated can persist in late times, or whether the final
equilibrium state is still entangled or not.

 At late times, the  two-atom subsystem will be in the asymptotic
equilibrium state. Though the effects of decoherence and
dissipation will generically make the state be separable so that
no entanglement is left in the end, there are also cases in which
the entanglement still exists at late times. To examine whether
the final equilibrium state is entangled or not, let us assume,
without loss of generality,  the reduced density matrix to have
the form
\begin{equation}\label{rho-1}
\rho(t)={1\over4}\bigg[\sigma_0\otimes\sigma_0+\rho_{0i}(t)\;\sigma_0\otimes\sigma_i
+\rho_{i0}(t)\sigma_i\otimes\sigma_0+\rho_{ij}(t)\sigma_i\otimes\sigma_j\bigg]\;,
\end{equation}
where the components $\rho_{0i}(t),\rho_{i0}(t),\rho_{ij}(t)$ are
real. Substituting Eq.~(\ref{rho-1}) into Eq.~(\ref{master}), we
can obtain, with setting $H_{\rm eff}=0$,
\begin{equation}\label{rho0i}
{\partial{\rho_{0i}(t)}\over\partial{t}}=-4A_1\rho_{0i}(t)-4B_1n_i-2B_2n_i
\tau+2B_2n_k\rho_{ik}(t)-2C_1\rho_{0i}(t)+2C_1n_in_k\rho_{0k}(t)\;,
\end{equation}
\begin{equation}\label{rhoi0}
{\partial{\rho_{i0}(t)}\over\partial{t}}=-4A_1\rho_{i0}(t)-4B_1n_i-2B_2n_i
\tau+2B_2n_k\rho_{ki}(t)-2C_1\rho_{i0}(t)+2C_1n_in_k\rho_{k0}(t)\;,
\end{equation}
\begin{eqnarray}\label{rhoij}
{\partial{\rho_{ij}(t)}\over\partial{t}}&=&-8A_1
\rho_{ij}(t)-4A_2\rho_{ji}(t)+4A_2\tau\delta_{ij}-4B_1[n_i\rho_{0j}(t)+n_j\rho_{i0}(t)]
\nonumber\\
&&-2B_2[n_i\rho_{j0}(t)+n_j\rho_{0i}(t)]+2B_2[n_k(\rho_{k0}(t)+\rho_{0k}(t))]\delta_{ij}\nonumber\\&&
-4 C_1\rho_{ij}(t)-4C_2\rho_{ji}(t)+4C_2[n_in_k\rho_{jk}(t)+n_jn_k\rho_{ki}(t)-n_in_j\tau]\nonumber\\
&&+ 2C_1[n_in_k\rho_{kj}(t)+n_jn_k\rho_{ik}(t)]
+4C_2[\tau-n_kn_l\rho_{kl}(t)]\delta_{ij}\;.
\end{eqnarray}
 Here, $\tau$ is the trace of the density matrix
 $\tau=\Sigma_{i=1}^3\rho_{ii}(t)$. Recall that $n_i$ are the
 components of the unit vector appearing, for example, in Eq.~(\ref{H-atom}) and
 Eq.~(\ref{psij}). If we symmetrize and anti-symmetrize the
 density matrix components $\rho_{oi}(t),\; \rho_{ij}(t)$, we can
 split the above system of differential equations into two
 independent sets. One can then show that the anti-symmetrized components
 decay exponentially as time grows while the symmetrized ones
 approach a non-zero asymptotic value. Therefore, there exists a
  final equilibrium state $\hat{\rho}$, the explicit form of which,  for a
  non-zero  atom separation, can be found by  setting  the right hand side of
  Eq.~(\ref{rho0i}),  Eq.~(\ref{rhoi0}),
  and Eq.~(\ref{rhoij}) to be zero, since any equilibrium state satisfies
  $\partial_t\hat{\rho}=0$, and the solution is
\begin{eqnarray}\label{density}
&&\tau={(2A_1+A_2)B_1(B_1-B_2)\over2{A_1}^3
-{A_1}^2A_2-A_2B_1B_2+A_1({B_2}^2-{A_2}^2)}\nonumber\\
&&\hat{\rho}_{0i}=\hat{\rho}_{i0}=-{({A_1-A_2})B_1(2A_1+A_2)n_i\over2{A_1}^3
-{A_1}^2A_2-A_2B_1B_2+A_1({B_2}^2-{A_2}^2)}\;, \nonumber \\
&&\hat{\rho}_{ij}={({A_1-A_2})B_1(2B_1+B_2)n_in_j\over2{A_1}^3
-{A_1}^2A_2-A_2B_1B_2+A_1({B_2}^2-{A_2}^2)}\;.
 \end{eqnarray}
To see if the equilibrium state is entangled or not, we will
calculate its concurrence which is defined to be $ {\cal C}[\rho]
=\max\{\lambda_1-\lambda_2 -\lambda_3-\lambda_4\}\;$, where
 $\lambda_{\mu}\;,\mu=1,2,3,4\;$, are the square roots of the
non-negative eigenvalues of the matrix $\rho\widetilde{\rho}$ in
decreasing order. Here, the auxiliary matrix
$\widetilde{\rho}=(\sigma_2\otimes\sigma_2)\rho^{T}(\sigma_2\otimes\sigma_2)\;$.
Substituting Eq.~(\ref{density}) into Eq.~(\ref{rho-1}), we can
easily calculate the concurrence for the unit vector
$\mathbf{{n}}$ along the third axis via the above the equations.
It is found that the concurrence is zero, which means that the
equilibrium state is separable and  entanglement generated
initially does not persist at late times. Same results can be
obtained for ${\bf {n}}$ along other directions.

However, it should be pointed out that all the expressions of
Eq.~(\ref{density}) become indefinite (of the form $0/0$) when the
 separation of the atom approaches zero which results in $A_1=A_2,B_1=B_2,C_1=C_2 $.  Therefore,
the case for the vanishing atom separation should be dealt with
separately. Taking the trace of both sides of Eq.~(\ref{rhoij})
for the vanishing atom separation, we find that
$\tau=\Sigma_{i=1}^3\rho_{ii}(t)$ is actually a constant of
motion, which is determined by the initial reduced density, while
the expression for $\tau$ in Eq.~(\ref{density}) is no longer
valid. In fact, the positivity of the initial density matrix
requires that $-3\leq\tau\leq 1$. Consequently, we should take
$\tau$ as a new independent parameter, and components of the
density matrix for the equilibrium state, $\hat{\rho}$, in the
present case, read
 \begin{eqnarray}
&&\hat{\rho}_{0i}=\hat{\rho}_{i0}=-{R\over 3+R^2}(\tau+3)n_i\;, \nonumber \\
&&\hat{\rho}_{ij}={1\over3+R^2}[(\tau-R^2)\delta_{ij}+R^2(\tau+3)n_in_j]\,,
 \end{eqnarray}
where $R=B_1/A_1$.   The corresponding concurrence can be
calculated directly
\begin{equation}
{\cal
C}(\hat{\rho})=\max\Bigg\{{(3-R^2)\over2(3+R^2)}\bigg[{5R^2-3\over3-R^2}-\tau\bigg],0\Bigg\}\;,
\end{equation}
which is non-zero provided $\tau$ for the initial state $\rho(0)$
obeys
\begin{equation}\label{ent-con}
\tau < {5R^2-3 \over 3-R^2}\;.
\end{equation}
This reveals that when the atom separation is zero ($L=0$), the
entanglement generated initially persists at late time despite of
the decoherence and dissipation of the external environment and
the late-time equilibrium state is still entangled, as long as
(\ref{ent-con}) holds. This is in sharp contrast with the case of
a non-zero separation. However, the presence of the boundary has
on effect on deciding whether the initially created entanglement
can be maintained at late times in the equilibrium state, since
 the concurrence is only dependent on $\tau$ and $R$, and
 factors containing the boundary parameter $z$ are all canceled
out in the expression of $R$ if one recalls Eq.~(\ref{abc}), thus
the concurrence for the final equilibrium state is independent of
the presence boundary.

\section{Discussion}

In summary,  we have examined the entanglement generation between
two mutually independent two-level atoms immersed in a thermal
bath of scalar particles subjected to a perfectly reflecting plane
boundary.  With the help of the master equation that describes the
evolution in time of the atom subsystem obtained by tracing over
environment (external scalar fields) degrees of freedom, we find
that the presence of the boundary may play a significant role in
controlling the entanglement creation in some circumstances and
the new parameter, the distance of the atoms from the boundary,
gives one more freedom in controlling the entanglement generation.

In particular,  when two atoms are placed very close to the
boundary, i.e., $z/L\ll 1$  and $\omega L$ is approximately less
than three, that is, when the separation, $L$, is approximately
less than three times the characteristic wavelength of the atom's
radiation, then for a certain temperature the presence of the
boundary will make the atoms be entangled which would otherwise
still be separable. Therefore the presence of the boundary gives
us more power in creating entanglement. However, when $\omega{L}$
is large, i.e., the separation is much larger than the
characteristic wavelength of the atom's radiation,  the presence
of the boundary will make the atoms less likely to be entangled
than otherwise.  Meanwhile  when $z/L$ is very large, i.e., when
the atoms are very far from the boundary,
 the influence of the presence
of the boundary on the entanglement generation is negligible as
expected.

At the same time, we find that the variation of  location of the
atoms has significant influence on entanglement generation between
two initially independent atom only when $\omega{L}$  is of order
one, or in different words,  both when $\omega{L}$ is very small
or very large, the variation of location of the atoms has no
appreciable effect on the entanglement generation. Note, however,
that this by no means suggests that the presence of the boundary
does not affect the entanglement generation.

Our analysis also reveals that  the entanglement generated because
of the correlations induced by the environment will persist in the
late time asymptotic equilibrium state if the separation between
the atoms is vanishing. However, when the separation is non-zero,
the entanglement will disappear at late times and the asymptotic
equilibrium state becomes unentangled again. Finally,  the
presence of a boundary generally has no effect on maintaining the
entanglement initially generated in the asymptotic equilibrium
state.

\begin{acknowledgments}
 This work was supported in part by the National
Natural Science Foundation of China  under Grants No. 10375023 and
No. 10575035, and the Program for New Century Excellent Talents in
University (NCET, No. 04-0784).
\end{acknowledgments}

\end{document}